appended ( 2 MO) can be obtained from the author.
\documentstyle[11pt]{spacekap}
\begin{opening}
\title{SYNTHESIS OF GALACTIC STELLAR POPULATIONS \\ AND EXPECTED
CONSTRAINTS FROM INFRARED SURVEYS}
\author{Annie C. Robin}
\institute{Observatoire de Besan\c{c}on, BP1615, F-25010
Besan\c{c}on Cedex, France}
\date{}
\end{opening}
\runningtitle{Infrared constraints on population synthesis}
\begin{document}
\begin{abstract}

Models of population synthesis for the Galaxy have been developed in order to
understand galactic structure and evolution. They allow to test scenarii of
evolution by comparisons between model predictions and observed distributions.
Forthcoming near-infrared surveys will enormously increase the amount of
information about the distributions of late type stars in the Galaxy. We show
here how population synthesis models will help to analyse these surveys and to
interprete them in terms of mass function, star formation history and stellar
evolution.

\end{abstract}

\section{Introduction}

First generation of models dedicated to understanding galactic structure was
generally limited to describe the present aspect of stellar populations.
For understanding more about the galaxy evolution
one needs to refere to knowledges about physical processes occuring in star
formation, dynamics and internal structure of stars and to link them to
observed properties and statistics.  The raw
observational data which can be obtained for remote stars do not allow
deriving intrinsic stellar parameters of individual stars, such as distance,
mass, age,
evolutionary stage, chemical composition, interstellar extinction.
However, some information relevant to the distribution
of these quantities is reflected in the n-dimensionnal distribution of
observables. Magnitudes and colours are also connected to ages and star
formation processes through the history of star formation and evolution.
A comprehensive model have been built in
Besan\c{c}on in order to extract as much informations as possible from present
distributions of stars and to interpret them in terms of galactic evolution
and structure.
We show how it can be used to analyse large infrared surveys and to get new
constraints on star formation history, mass function of faint stars and late
stages of stellar evolution.

\section{Models of population synthesis}

\subsection{Models describing galactic structure}

In the visible, the first generation of galaxy models dedicated to
predict and interpret star counts were based on the equation of stellar
statistics (van Rhijn, 1965). Assuming density laws
and luminosity functions for stellar populations, they estimate the
number of stars integrated along a line of sight. These models
are generally able to fit star counts in magnitude and wide band colour
(see Robin, 1993 for references) but
the solution is not unique, as can be seen on the results obtained by
different authors for galactic structure parameters.

In the near-infrared a set
of models of the same type have been developed, generally adding to the
population in the visible populations
contributing mainly in the infrared, like AGB's, and eventually spiral
arms and a molecular ring (Jones et al., 1981, Garwood \& Jones, 1987,
Ruelas-Mayorga, 1991, Wainscoat et al., 1992, Ortiz \& Lepine, 1993).

\subsection{A model for understanding evolution}

To go further in understanding the history of our Galaxy, it is
necessary to account for the physical links between the density,
velocity and metallicity distributions in the Galaxy. Theory of
galactic dynamics and evolution, as well as the
observations show that the kinematics, the scale heights and the
metallicities are related to the age of the population. Such that a
reasonnable way to model the Galaxy is to use the age as the
basic parameter. The age distribution of a population can be predicted
by a scenario of galactic and stellar evolution. This approach has the
great advantage to allow to check the consistency between our
knowledges of stellar evolution, galactic, chemical and dynamical
evolution. Such model can start from an evolution scenario (IMF, SFR), a
set of evolutionary tracks, a galactic potential, a chemical evolution
scenario and allow to compute the present day stellar distribution all
over the Galaxy, on the point of view of the photometry, astrometry,
radial velocities and metallicities. Such a policy has been used by the
Besancon group to built a
self-consistent galaxy model (Robin \& Cr\'ez\'e, 1986), Bienaym\'e et al.,
1987, Haywood, 1993). The
usefullness of this approach has been proved by the numerous results
concerning galactic structure and evolution, obtained from photometric
and astrometric star counts in the visible (Robin, 1993).

By far this kind of model has less
free parameters than any empirical models because parameters are
self-consistently forced to follow physical laws. The limitation of
this approach is the difficulty to handle the multivariate distributions
of observational parameters which are needed to constrain the
hypotheses.

\subsection{Extension to infrared bands}

The extension of the Besan\c{c}on model to the infrared bands has been
made simply by using empirical calibrations for the J, H, K and L bands,
using Koornneef (1983) calibrations. Late type giants were added to the
model, since no evolutionary tracks ran until the AGB's. The
calibrations in absolute magnitude for carbon stars and OH/IR has been made by
Guglielmo (1993, and this conference). Figure~1
shows model predictions
in a set of directions in the K band compared to observations. The
agreement is very good, except at longitude 320\deg~ corresponding to a spiral
arm.

\begin{figure}
\vspace{5cm}
\caption{Model predictions (solid lines) in 6 directions compared to data
from Elias (squares) and Jones et al. (diamonds)}\label{pred}
\end{figure}

\section{Constraints on evolution expected from IR surveys}

Specificity of the infrared spectral domain gives the opportunity to
observe objects at large distances in the galactic plane, thanks to the
lower extinction at 2 microns. At these wavelengths one are able to study the
distribution of population I stars in a wide part of the galaxy.
Bright stars in the infrared, supergiants and late type giants,
will be observable all over the Galaxy (see Jura this conference). They
can be used as tracers of the large scale structure.

On the contrary
very low mass stars would be detectable in a small volume around the
sun. But we expect to observe a sufficient number of them to obtain good
constraints on their luminosity function. From a survey in K to magnitude 14
on one hemisphere we expect to observe between 2000 and 4000 stars fainter than
12 in M$_{bol}$.

\subsection{Constraints on evolution scenarios}

Star counts are sensitive to the age distribution (then to the star
formation rate) since, because of orbit diffusions in the disc, the scale
heights vary with time. The dwarf to giant ratios are related to the
initial mass function and the variation of the star formation rate in
the past. Some colours are also slightly sensitive to the age distribution
because the metallicities have varied with time. This can be seen in some
examples given below.

\subsubsection{Evolution of the disc}

Figure~2 shows the distributions of stars in the (K, I-K) diagram
towards
the anticentre at low latitude (b=3), with young stars in (a) and old stars
in (b). The red part of the diagram is
dominated by giants while the blue part contents main sequence stars.
Changing the emphasized SFR history in the model would lead to
significant changes in this diagram, specially the proportion
of giant to dwarfs. The same diagrams but at slightly higher latitude (b
=20) are given in fig. \label{anti2}. In this case, all giants at
magnitude less than 12 are old, because young ones are closer to the
plane. So analysing the ratio of giant to main sequence stars
at different apparent magnitudes and at different latitudes will allow to
constrain the age distribution, hence the variation of the SFR
with time.

\begin{figure}
\vspace{5cm}
\caption{Distribution of stars in the (K,I-K) diagram according to their
age. (a) stars younger than 3 Gyr. (b) stars older than 3 gyr. Towards
(180,3)}\label{anti1}
\end{figure}
\begin{figure}
\vspace{5cm}
\caption{Distribution of stars in the (K,I-K) diagram according to their
age. (a) stars younger than 3 Gyr. (b) stars older than 3 gyr. Towards
(180,20)}\label{anti2}
\end{figure}
\begin{figure}
\vspace{5cm}
\caption{Distribution of stars in the (K,J-K) diagram towards (180,3).
(a) stars closer than
5.5 kpc. (b) Stars farther than 5.5 kpc.}\label{cut}
\end{figure}

\subsubsection{Disc structure}

As recently shown by Robin et al. (1992) the disc in the visible has a
scale length of 2.5 $\pm$ 0.3 kpc and appears to have a sharp cutoff at
$\approx 5.5$ kpc from the sun. These values may be measured using
near-infrared surveys, using slightly different populations. Figure~4
shows the appearance of such a cutoff in a (K, J-K) diagram
towards the anticenter. The cutoff will be clearly visible at 12$\le$K$\le$14
and 1.5$\le$J-K$\le$2 if the stellar populations are the same as in the
visible.
If different populations have different scale lengths and cutoffs (as
suggested for carbon stars and oxygen stars) it
will be of first importance for galactic studies to measured these values
for different samples in the infrared surveys.

Carney \& Seitzer (1993) detected the warp followed by stars in
the direction l=240\deg and l=270\deg, confirmed by the Dirbe experiment.
The analysis of the stellar distribution in this remote part of the disc
will give new constraints on the large scale potential of the Galaxy.

\subsubsection{The bulge population}

The bulge is certainly one of the most interesting region to explore with
a near-infrared survey. The relatively low extinction in the K bands will allow
to observe close to the center of the Galaxy. However one expects to be limited
by the confusion of the sources in the K band at a longitude smaller than
30\deg in the
plane and at latitude smaller that 5\deg at longitude 0\deg.
This will however allow to study the overall shape of the bulge (peanut-shape,
disk-like or bar) and its link with the disc, thick disc and halo.

\subsubsection{Thick disc and halo populations}

The expected insight on these old populations from IR surveys is merely
not different from visible surveys.
The sample of known halo giants is presently restricted to less than
few hundreds of objects. In this case the luminous part of the mass function
is very poorly known and assumed to be like the one of globular clusters. Wide
field surveys will help to check this hypothesis. On the faint end of the
mass function also, constraints can be derived from infrared surveys. However
this will be more difficult than in the disc because of the scarcity of
population II stars in the solar neighbourhood and the planned surveys do not
go to
faint enough magnitudes for getting a sufficient number of very low mass halo
stars. The constraints would come on intermediate mass stars. An overall
analysis of numerous field at the same time will be needed in order to get
these luminosity and mass functions.

\section{Conclusions}

Infrared surveys are expected to give new constraints on galactic evolution.
A large part of the advancement coming from those will benefit from cross
identification with the visible and far infrared surveys.
However the analysis of the whole data sets will be difficult to
handle. Methods of statistical analysis have to be developed specially for
large surveys. These methods should be multivariate and able to get most of
the information on a systematic way. Models of population synthesis will be
helpful tools for this part of the work. Methods like density estimation (
Chen, 1993), hierarchical or non hierarchical classifications would help when
adapted for the analysis of forthcoming large digitised surveys.
Meanwhile the huge data sets available from near-infrared surveys will give
new constraints on late stage of stellar evolution, galactic structure and
evolution.

\end{document}